\documentclass[12pt]{article}
\usepackage{amsmath,latexsym,amssymb}
\newtheorem{theorem}{Theorem}

\def\cH{{\cal H}}
\def\Tr{\mathop{\rm Tr}\nolimits}

\def\rank{\mathop{\rm rank}\nolimits}

\begin{document}
\title{Simple construction of\\
quantum universal variable-length source coding}
\author{
Masahito Hayashi
\thanks{Laboratory for Mathematical Neuroscience, 
Brain Science Institute, RIKEN,
2-1 Hirosawa, Wako, Saitama, 351-0198, Japan}
 and Keiji Matsumoto\thanks{
Quantum Computation and Information Project, ERATO, JST,
5-28-3, Hongo, Bunkyo-ku, Tokyo, 113-0033, Japan}}
\date{}
\maketitle
\begin{abstract}
We simply construct a {\it quantum universal variable-length source code}
in which,
independent of information source, both of the average error and 
the probability that the coding rate is 
greater than the entropy rate $H(\overline{\rho}_p)$, tend to $0$.
If $H(\overline{\rho}_p)$ is estimated, we can compress
the coding rate to the admissible rate
$H(\overline{\rho}_p)$ with a probability close to 1.
However, when we perform a naive measurement for the
estimation of $H(\overline{\rho}_p)$,
the input state is demolished.
By smearing the measurement,
we successfully treat the trade-off
between the estimation of $H(\overline{\rho}_p)$
and the non-demolition of the input state.
Our protocol can be used not only for the Schumacher's scheme
but also for the compression of entangled states.
\end{abstract}
\section{Introduction}	        
When we compress our data in a classical computer,
we usually use not a fixed-length code, but
a variable-length source code like gzip.
In the quantum case,
according to Schumacher's results\cite{Schumacher},
when our quantum data obeys independent identical distribution
(i.i.d.) of a probability $p$ of quantum states,
we can compress our data up to the entropy rate of 
the average density operator defined as the mixture of 
the probability $p$.
However, his protocol is not applicable 
to the case where we do not know the average density operator,
which the construction of the protocol is dependent on.
Using  representation theory of unitary group,
Jozsa and Horodecki family\cite{JH} constructed
a quantum universal fixed-length code,
and it is efficient in the i.i.d. case 
when the entropy rate of the source is less than the rate of the code.
Otherwise, this protocol demolish the state unrecoverablly.
(The optimality of their code among  quantum fixed-length codes
is proven not only in the sense of the compression rate, but also
in the sense of error exponent by Hayashi\cite{0202002}.)

Hence, a quantum universal variable-length code,
that does not depend on the rate is more desired.
Of course, in such a code,
the coding rate must not be determined a priori 
and it must be decided from the input state.
While this decision does not change
the source in the classical case, 
it does cause the destruction in the quantum case
because this decision requires a quantum measurement.
Therefore, we treat the trade-off between the compression rate
and the non-demolition.
While this type code was thought to be impossible by some researchers\cite{BF},
it was constructed by Hayashi and Matsumoto\cite{0202001}
as the following strategy.

First, we consider the optimal measurement
for the decision of the coding length (the estimation of the
entropy rate) in the sense of the large deviation
(also  optimal in the sense of mean square error, in many cases).
For such an optimal measurement demolishes the state,
an unsharp measurement, generated by smearing out the optimal measurement,
is considered.
Such an unsharp one is also optimal in the sense of the large deviation
while it is not optimal in the sense of the mean square error any more.
The previous paper\cite{0202001} used  such a smeared measurement,
but the smearing process is too complicated to give a clear insight
to the essence of the protocol.
In this paper, 
by constructing a quantum variable-length code from 
a quantum universal fixed-length code,
we clarify the 
trade-off between the compression rate
and the non-demolition.

To construct optimal code
in the sense of overflow exponent,
we make heavy use of the group representation
type theory, which is not necessary 
to  achieve the optimal compression rate.
For example, one can replace our group representation theoretic
estimation of the entropy of the average state 
by the one based on a tomographic estimate of 
the state. In this case, evaluation of the protocol will be done 
by usual type theory (no room for group representation!).

Based on the idea of smeared measurement 
of Hayashi and Matsumoto\cite{0202001},
Presnell and Jozsa\cite{PJ,PJ2} developed the following strategy.
We decide the coding length for every 
computational basis and compress the input state
by using a classical compression algorithm
under the computational basis
having the minimum coding length.
Since the demolition of the state
is unavoidable in their approach, 
they applied the above smearing method to their approach, and 
successfully constructed 
a quantum universal variable-length source code.
Their code achieves optimal compression rate not only for 
the i.i.d. case but also for other cases.
For example, when 
states generated from the source is orthogonal with each other,
their method is useful though the probability 
is not i.i.d. but Markov.
However, their method has the following drawbacks.
While this paper and the paper\cite{0202001}
optimize the decreasing exponent of the overflow probability,
their paper did not deal with this type optimization.

The essential point of this paper is 
the construction of a quantum universal variable-length code
from a quantum universal fixed-length code
which satisfies the large deviation principle.
In the present protocol, 
when quantum information sources 
generate non-orthogonal states 
with a classical Markov chain probability,
it seems difficult to achieve the optimal compression rate.
(Of course it seems difficult also in Presnell and Jozsa's protocol.)
However, if we successfully construct a quantum universal fixed-length code
for this case which satisfies the large deviation principle,
we seem able to construct a quantum universal variable-length code
for this case by modifying the quantum universal fixed-length code.

This paper is organized as follows.
First, we review quantum fixed-length source code
which contains quantum universal fixed-length source code.
Second, we give a precisely definition of 
quantum universal variable-length source code.
Next, we construct a more simple code from
a quantum universal fixed-length code.
Finally, we discuss a application of our protocol to
a compression of entangled states.

\section{Review of quantum fixed-length source coding}\label{s2.1}
Let $\cH$ be a finite-dimensional Hilbert space 
that represents the physical system of interest
and let ${\cal S}(\cH)$ be the set of
density operators on $\cH$.
Consider a source which produces
the 
state $\vec{\rho}_n
:=\rho_1 \otimes \rho_2 \otimes \cdots \otimes\rho_n$
with the i.i.d. distribution $p^n$
of the probability $p$ on 
states.
In {\it fixed-length source coding},
a sequence of states 
$\vec{\rho}_n$
is compressed 
to the state in a smaller Hilbert space 
$\cH_n \subset \cH^{\otimes n}$,
whose dimension is $e^{nR}$.
Here, the encoder and the decoder is
a trace-preserving completely positive 
(TP-CP) map $E^n$ and $D^n$, respectively. 
The average of the total error is given by
\begin{align*}
\epsilon_{n,p}(E^n, D^n):=
\sum_{\vec{\rho}_n \in {\cal S}(\cH^{\otimes n})}
p^n(\vec{\rho}_n)
b^2\left(\vec{\rho}_n,  
D^n \circ E^n( \vec{\rho}_n)
\right) , 
\end{align*}
where
Bures' distance is defined as
$b(\rho,\sigma):= \sqrt{
1- \Tr \left|
\sqrt{\rho}\sqrt{\sigma}\right|}$.
In this setting, we focus on the infimum of the rate 
with which the average error goes to zero.
The infimum is called the minimum admissible rate $R_p$ 
of $p$, and is defined by
\begin{align*}
R_p 
:=
\inf\left\{\left.
\limsup \frac{1}{n} \log \dim \cH_n \right|
\begin{array}{c}
\exists \{ (\cH_n, E^n,D^n) \}, \\
\epsilon_{n,p}(E^n, D^n) \to 0 
\end{array}
\right\}.
\end{align*}
As was proven by 
Schumacher \cite{Schumacher}, and Jozsa and Schumacher
\cite{JS}, and Barnum et al. \cite{Barnum},
when every $\rho_i$ is pure, 
the equation 
$R_p= H( \overline{\rho}_p):= 
-\Tr \overline{\rho}_p\log \overline{\rho}_p $ holds,
where $\overline{\rho}_p:=
\sum_{\rho \in {\cal S}(\cH)}
p(\rho) \rho$.
Moreover Jozsa et al. \cite{JH}
constructed the projections $P_{R,n}$ 
for a arbitrary rate $R$ such that
\begin{align}
\frac{1}{n}\log \rank P_{R,n} \to R,
\quad
\Tr  P_{R,n} \rho^{\otimes n} \to 1, \label{E-3}
\end{align}
for any density matrix $\rho$ satisfying $H(\rho) \,< R$.
Using the above projections, they 
proposed a {\it quantum universal fixed-length source code}
depending only on the entropy rate as follows.
The encoder $E^n$ is defined by
\begin{align*}
E^n_R(\vec{\rho_n}):=
P_{R,n}\vec{\rho_n}P_{R,n} +
\left(
\Tr (I- P_{R,n})\vec{\rho_n} \right)
|0 \rangle \langle 0|,
\end{align*}
and the decoder $D^n_R$ is defined as the embedding.

Hayashi more precisely evaluated the the performance of
their code as follows\cite{0202002}.
We can chose a projection $P_{R,n}$ such that
\begin{align}
\rank P_{R,n} \le (n+1)^{2d}(n+d)^{2d}e^{n R} \le
(n+d)^{4d}e^{n R} \label{10-1} \\
1- \Tr  P_{R,n} \overline{\rho}_p^{\otimes n} 
\le
(n+d)^{4d}
\exp\left(- n \min_{H({\bf b})\ge R}
D({\bf b}\|{\bf a}) \right),\label{10-2} 
\end{align}
where ${\bf a}$ denotes the probability distribution 
consisting of the eigenvalues of $\overline{\rho}_p$
and ${\bf b}$ denotes another probability distribution.
Therefore, the average of the total error
is evaluated by
\begin{align}
\epsilon (E^n_{R},D^n_{R})
\le 2
 (n+d)^{4d}
\exp\left(- n \min_{H({\bf b})\ge R} 
D({\bf b}\|{\bf a})\right), \label{19-7}
\end{align}
which goes to $0$ when $R \,< H(\overline{\rho}_p)$.
The inequality (\ref{19-7})
was proven in the pure state case in Hayashi \cite{0202002},
but as is proven in Appendix,
it holds in the mixed state case.
This type evaluation was essentially done by Keyl and Werner\cite{KW}.

\section{Quantum universal variable-length source coding}
\label{s2.2}
In the classical system,
depending on the input state,
the encoder can determine the coding length.
Such a code is called a variable-length code.
Using this type of code, we can compress any information 
without error.
Lynch \cite{Ly} and Davisson \cite{Da}
proposed a variable-length code with no error,
in which the coding rate is less than
$H(p)$ except for a small enough probability under 
the distribution $p$.
Such a code is called
a universal variable-length source code.
Today, their code can be regarded 
as the following two-stage code:
at the first step, we send the empirical distribution 
which indicates a subset of data,
and in the second step, we send information
which indicates every sequence belonging to the subset.

This paper deals with 
quantum data compression 
in which the encoder determines the coding length, 
according to the input state.
In order to make this decision, he must measure the input 
quantum system.
As is known, any quantum measurement is described by
POVM $M= \{M_\omega\}_{\omega \in \Omega}$.
When the data set $\Omega$ is discrete,
we may describe the state evolution of 
a quantum measurement $M$ as follows
while 
we need an instrument, i.e., CP-map valued measure
in the general case\cite{Oz1}.
When we perform a quantum measurement $M= \{M_\omega\}_{\omega 
\in \Omega}$
to the quantum system whose state is a density operator $\rho$,
we obtain the data $\omega \in \Omega$ with the probability 
$\Tr \rho M_\omega$ and 
the final state is
$\sqrt{M_\omega} \rho \sqrt{M_\omega}/
\Tr \rho M_\omega$.
An encoding process after the measurement
is described by a TP-CP map $E_{\omega}$.
Therefore, any encoder is given
by ${\bf E}_{\omega}(\rho):=E_{\omega}
(\sqrt{M_\omega} \rho \sqrt{M_\omega})$,
~(${\bf E}=\{{\bf E}_{\omega}\}_{\omega \in \Omega}$).
The decoder is given by 
a set of TP-CP maps ${\bf D} = \{ {\bf D}_{\omega}\}_{
\omega \in \Omega}$,
which presents the decoding process depending on
the data $\omega$.
A pair of an encoder ${\bf E} = \{ 
{\bf E}_{\omega} \}_{\omega \in \Omega}$ and 
a decoder ${\bf D} = \{ {\bf D}_{\omega}\}_{
\omega \in \Omega}$
is called 
a {\it quantum variable-length source code}
on $\cH$.
The coding length is described by
$\log | \Omega | + \log \dim \cH_{\omega}$,
which is a random variable
obeying the probability
${\rm P}_{\rho}^{{\bf E}}(\omega)
:=
\Tr {\bf E}_{\omega}(\rho)$
when the input state is $\rho$.

When the state $\vec{\rho}_n$ on 
$\cH^{\otimes n}$
obeys the i.i.d. 
distribution $p^n$ of the probability
$p$ on 
states,
the error of decoding for 
a variable-length code $({\bf E}^n,{\bf D}^n)$
on $\cH^{\otimes n}$
is
evaluated by Bures' distance as 
\begin{align*}
\sum_{\omega_n \in \Omega_n}
\Tr {\bf E}^n_{\omega_n}(\vec{\rho}_n)
b^2\left(\vec{\rho}_n,  
{\bf D}^n_{\omega_n} \left( 
\frac{{\bf E}^n_{\omega_n}(\vec{\rho}_n)}
{\Tr {\bf E}^n_{\omega_n}(\vec{\rho}_n)}
\right) \right), 
\end{align*}
and 
the average error is given by
\begin{align*}
\epsilon_{n,p}({\bf E}^n,{\bf D}^n) := 
\sum_{\vec{\rho}_n \in {\cal S}(\cH^{\otimes n})}
p^n(\vec{\rho}_n)  \sum_{\omega_n \in \Omega_n}
\Tr {\bf E}^n_{\omega_n}(\vec{\rho}_n)
b^2\left(\vec{\rho}_n,  
{\bf D}^n_{\omega_n} \left( 
\frac{{\bf E}^n_{\omega_n}(\vec{\rho}_n)}
{\Tr {\bf E}^n_{\omega_n}(\vec{\rho}_n)}
\right) \right).
\end{align*}
In this case, 
the data $\omega_n$ obeys the probability:
\begin{align*}
{\rm P}_{n,p}^{{\bf E}^n}(\omega_n)
:=
\sum_{\vec{\rho}_n \in {\cal S}(\cH^{\otimes n})}
p^n(\vec{\rho}_n)
\Tr {\bf E}^n_{\omega_n}(\vec{\rho}_n)
=
\Tr {\bf E}^n_{\omega_n}(\overline{\rho}_p^{\otimes n})
. 
\end{align*}
A sequence $\{({\bf E}^n,{\bf D}^n)\}$
of 
quantum variable-length source code
is called {\it universal}
if 
$\epsilon_{n,p}({\bf E}^n,{\bf D}^n) \to 0 $
for any probability $p$ on 
states.

As mentioned latter, 
there exists a quantum universal 
variable-length source code
$\{({\bf E}^{n}, {\bf D}^{n})\}$ satisfying
\begin{align*}
\lim
{\rm P}_{n,p}^{{\bf E}^n}
\left\{ \frac{1}{n}
\left( \log | \Omega_n | + \log \dim \cH_{\omega_n}
\right) \ge H(\overline{\rho}_p) +\epsilon
 \right\}
 = 0
\end{align*}
for any $\epsilon \,> 0$.
Conversely, if 
a quantum variable-length source code
$\{({\bf E}^{n}, {\bf D}^{n})\}$
is universal and
\begin{align*}
\lim
{\rm P}_{n,p}^{{\bf E}^n}
\left\{ \frac{1}{n}
\left( \log | \Omega_n | + \log \dim \cH_{\omega_n}
\right) \ge R
 \right\}
= 0, 
\end{align*}
then $R \ge R_p= H(\overline{\rho}_p)$.
Moreover, concerning the exponent of the overflow probability
${\rm P}_{n,p}^{{\bf E}^n}
\left\{ \frac{1}{n}
\left( \log | \Omega_n | + \log \dim \cH_{\omega_n}
\right)\ge R \right\}$
the following theorems hold\cite{0202001}.
 
\begin{theorem}\label{t2}
There exists a quantum universal variable-length source code
$\{({\bf E}^{n}, {\bf D}^{n})\}$ 
on $\cH^{\otimes n}$ such that
\begin{align}
\lim
\frac{-1}{n}\log 
{\rm P}_{n,p}^{{\bf E}^n}
\left\{ \frac{1}{n}
\left( \log | \Omega_n | + \log \dim \cH_{\omega_n}
\right) \ge R \right\}
=
\inf_{{\bf b}: H({\bf b}) \ge R}
D({\bf b} \| {\bf a}), 
\label{op3}
\end{align}
where ${\bf a}$ is a probability distribution consisting of
the eigenvalues of $\overline{\rho}_p$,
and ${\bf b}$ denotes another probability distribution.
$D({\bf b}\|{\bf a})$ 
is relative entropy.
\end{theorem}

\begin{theorem}\label{t1}
If a sequence $\{({\bf E}^n,{\bf D}^n)\}$
of quantum variable-length source codes
on $\cH^{\otimes n}$
is universal,
then
\begin{align}
\limsup
\frac{-1}{n}\log 
{\rm P}_{n,p}^{{\bf E}^n}
\left\{ \frac{1}{n}
\left( \log | \Omega_n | + \log \dim \cH_{\omega_n}
\right) \ge R \right\}
\le 
\inf_{{\bf b}: H({\bf b}) \ge R}
D({\bf b} \| {\bf a})
\end{align}
\end{theorem}
Therefore, the RHS of (\ref{op3}) is the optimal 
exponent of the overflow probability.

\section{Construction of a 
quantum variable-length source code}\label{s4}
First, for an intuitive explanation of
our construction, 
we naively construct a good variable-length code.
For this construction,
we fixed a strictly increasing sequence 
$\vec{\alpha}:=\{ \alpha_i\}_{i=1}^{l+1}$
of real numbers
such that
$0 =\alpha_1 \,< \alpha_2 \,< \ldots \,< \alpha_l \,< \alpha_{l+1}= \log d$.
We define the encoder ${\bf E}^{\vec{\alpha},n}$
with the data set $\{ 1, \ldots, l\}$ by
\begin{align*}
P^{\vec{\alpha},n}_i &:= P_{\alpha_{i+1},n}- P_{\alpha_{i},n}  \\
{\bf E}_i^{\vec{\alpha},n}(\rho_n)
&:=
P^{\vec{\alpha},n}_i  \rho_n P^{\vec{\alpha},n}_i ,\quad
\rho_n \in {\cal S}(\cH^{\otimes n}),
\end{align*}
and define the decoder ${\bf D}_i^{\vec{\alpha},n}$
as the embedding to $\cH^{\otimes n}$.
Assume that $H(\overline{\rho}_p)$
belongs to the interval $[  \alpha_i,\alpha_{i+1})$.
As is guaranteed by (\ref{E-3}),
if $H(\overline{\rho}_p)$ does not
lie on the boundary on the open interval $(  \alpha_i,\alpha_{i+1})$,
the probability $\Tr \overline{\rho}_p^{\otimes n}
P^{\vec{\alpha},n}_i $ tends to $1$.
Thus, we can prove $\epsilon_{n,p}(
{\bf E}^{\vec{\alpha},n},{\bf D}^{\vec{\alpha},n})  \to 0$.
Of course, if we choose $\alpha_{i+1}-\alpha_i$ to be sufficiently
small,
the coding length is close to the entropy 
$H(\overline{\rho}_p)$
with almost probability $1$.
However, if $H(\overline{\rho}_p)$ lies on 
the boundary, i.e. $H(\overline{\rho}_p)= \alpha_i$,
the state is demolished,
as is caused by the same reason of Lemma 2 in \cite{0202001}.
In this case, we can prove
$\lim \epsilon_{n,p}(
{\bf E}^{\vec{\alpha},n},{\bf D}^{\vec{\alpha},n})  \,> 0$.
Thus, it is not universal.

For the non-demolition of initial states,
we construct a variable-length code,
by choosing the integer $k$ such that $0\,< k \le \delta n$
at random
where $\delta:= \log d/ (l-1)$.
Depending on the integer $k$, we define 
$\vec{\alpha}(k/n)=\{ \alpha(k/n)_i\}_{i=1}^{l+1}$ as
\begin{align*}
\alpha(k/n)_i &= k/n + (i-2)\delta \hbox{ if }i=2, \ldots l. \\
\alpha(k/n)_1 &= 0, \quad \alpha(k/n)_{l+1} = \log d ,
\end{align*}
and use the encoder ${\bf E}^{\vec{\alpha}(k/n),n}$ and 
the decoder ${\bf D}^{\vec{\alpha}(k/n),n}$.
In this protocol,
the probability that $H(\overline{\rho}_p)$ lies on 
the boundary goes to zero.
Therefore, the above code seems a quantum universal
variable-length code.

In the following, we give a mathematical definition the above code,
evaluate its performance, 
and prove the optimality of
its optimality of the exponent of the overflow probability
by choosing $\delta$ (or $l$) depending on $n$.
We define the data set and the encoder ${\bf E}^{\delta,n}$
as
\begin{align*}
\Omega_n&:= 
\{ k \in\mathbb{Z} | 0 \,< k \le \delta n\}\times \{1, 2, \ldots, l\},
\\
{\bf E}^{\delta,n}_{k,i}&:= \frac{1}{[n\delta]}
{\bf E}^{\vec{\alpha}(k/n),n}_i ,
\end{align*}
and the decoder ${\bf D}^{\delta,n}_{k,i}$ as the embedding,
i.e., we perform the measurement 
\begin{align*}
\left\{\frac{1}{[n \delta]} P_i^{\vec{\alpha}(k/n), n}\right\}_{k,i}
\end{align*}
in the encoding process, where
$[x]$ is Gauss notation i.e., $[x]$ is the maximum integer $n$
satisfying $n \le x$.

Its performance was evaluated as follows:
\begin{align}
\epsilon_{n,p}({\bf E}^{\delta,n},{\bf D}^{\delta,n})
& \le
1- \frac{[n(\delta-2 \delta')]}{[n \delta]}
\left(1- \left(n+d\right)^{4d}
\exp(-n C {\delta'}^2 )
\right)^{3/2} \label{9-5-1} \\
{\rm P}^{{\bf E}^{\delta,n}}_{n,p}
\left\{
\frac{1}{n}
\left( \log | \Omega_n |+ \log
\dim \cH_{k,i}\right)
\ge R
\right\}
& \le
(n+d)^{4d}
\exp\left(- n \min_{H({\bf b})\ge R - f(n,\delta)/n}
D({\bf b}\|{\bf a}) \right) \label{9-5-2} \\
C&:= \min_{{\bf b}} \frac{D({\bf b}\|{\bf a})}{|H({\bf a})
- H({\bf b})|^2}
\label{103} \\
f(n,\delta)& := \delta + \log \frac{(n+d)^{4d}}
{[n \delta] ((\log d)/ \delta+1) }, \nonumber
\end{align}
where $\delta'$ is arbitrary real number satisfying $0 \,<2 \delta' \,<
\delta$.
The above inequalities are proven in Appendix.
When we choose
$\delta$ and $\delta'$ as
$\delta_n:= (1/n)^{1/6}$ and $\delta_n':= (1/n)^{1/3}$,
$f(n,\delta_n)/n$ goes to $0$.
Thus, we obtain 
\begin{align}
\epsilon_{n,p}({\bf E}^{\delta_n,n},{\bf D}^{\delta,n})
& \to 0 \label{19-9}\\
\liminf
\frac{-1}{n}\log 
{\rm P}^{{\bf E}^{\delta,n}}_{n,p}
\left\{
\frac{1}{n}
\left( \log | \Omega_n |+ \log
\dim \cH_{k,i}\right)
\ge R
\right\}
& \ge
\inf_{{\bf b}: H({\bf b}) \ge R}
D({\bf b} \| {\bf a}), \label{19-10}
\end{align}
which imply Theorem \ref{t2}.

\section{Compression of entangled states}\label{s5}
Next, we consider another compression problem
in which we compress an entangled state by local operations.
We apply our protocol given in section 4 to this problem.
Assume that we share an 
entangled state $|\phi \rangle \langle \phi|^{\otimes n}$
which is the tensor product of a pure state 
$|\phi \rangle \langle \phi|$ on the composite
system $\cH_A \otimes \cH_B$.
We want to save the dimension of the quantum system of Alice,
in the situation where only local operations of Alice's system
are allowed.
In a fixed-length compression, 
our operation is described by 
a triplet $(\cH_n, E^n,D^n)$ consisting of 
a subspace $\cH_n$ of $\cH_A^{\otimes n}$,
an encoder, i.e., a TP-CP map from $\cH_A^{\otimes n}$ to $\cH_n$,
and a decoder i.e., a TP-CP map from $\cH_n$ to $\cH_A^{\otimes n}$.
Its performance is characterized by the coding length 
$\log \dim \cH_n$ and the Bures' distance 
$b^2(|\phi \rangle \langle \phi|^{\otimes n},
D^n \circ E^n \otimes I_B^{\otimes n} 
(|\phi \rangle \langle \phi|^{\otimes n}))$,
where $I_B$ is the identity operator on $\cH_B$.
Now, we define the following value as the bound of the asymptotic performance:
\begin{align*}
R_{|\phi \rangle \langle \phi|}: =
\inf \left\{ \left. \limsup \frac{1}{n} \log \dim \cH_n \right|
\begin{array}{l}
\exists \{(\cH_n, E^n,D^n) \}, \\
b^2(|\phi \rangle \langle \phi|^{\otimes n},
D^n \circ E^n (|\phi \rangle \langle \phi|^{\otimes n})) \to 0 
\end{array}
\right\},
\end{align*}
then we can prove 
$R_{|\phi \rangle \langle \phi|}= H(\Tr_B |\phi \rangle \langle \phi|)$ 
as follows.
If $R \,> H(\Tr_B |\phi \rangle \langle \phi|)$,
then 
$\Tr (P_{R,n} \otimes I_B^{\otimes n}  ) 
|\phi \rangle \langle \phi|^{\otimes n}  = 
\Tr_A P_{R,n} \Tr_B |\phi \rangle \langle \phi|^{\otimes n} \to 1$.
Therefore, the encoder $E^n_R$ and the decoder $D^n_R$
defined in section 2
satisfies 
$b^2(|\phi \rangle \langle \phi|^{\otimes n},
D^n_R \circ E^n_R \otimes I_B^{\otimes n} (|\phi \rangle \langle
\phi|^{\otimes n})) \to 0$.
Conversely, we assume that a sequence $\{(\cH_n, E^n,D^n) \}$ satisfies
\begin{align}
b^2(|\phi \rangle \langle \phi|^{\otimes n},
D^n \circ E^n \otimes I_B^{\otimes n} (|\phi \rangle \langle
\phi|^{\otimes n})) \to 0 \label{19-2}.
\end{align}
The entanglement of formation of the compressed state
$E_f(E^n \otimes I_B^{\otimes n} (|\phi \rangle \langle
\phi|^{\otimes n}))$
is less than
$\log \dim \cH_n$, and 
the entanglement of formation of the original state
$E_f(|\phi \rangle \langle\phi|^{\otimes n}$) 
equals $n H(\Tr_B |\phi \rangle \langle \phi|) $.
(Concerning the entanglement of formation, please see, for example, 
Hayden et.al.\cite{Hayd}.)
Thus, similarly to Hayden et.al.\cite{Hayd}, 
by using (\ref{19-2}) and the continuity and the
monotonicity of the entanglement of formation, 
we can prove
\begin{align*}
& \limsup \frac{1}{n} \log \dim \cH_n \ge
 \limsup \frac{1}{n} E_f(E^n \otimes I_B^{\otimes n} (|\phi \rangle \langle
\phi|^{\otimes n})) \\
 \ge & \limsup \frac{1}{n} E_f(
D^n \circ E^n \otimes I_B^{\otimes n} (|\phi \rangle \langle
\phi|^{\otimes n}))) 
= \limsup \frac{1}{n} 
E_f(|\phi \rangle \langle\phi|^{\otimes n}) =
H(\Tr_B |\phi \rangle \langle \phi|) .
\end{align*}

However, the above protocol cannot be used when 
the entropy rate $H(\Tr_B |\phi \rangle \langle \phi|) $
is unknown.
In the following,
we consider the case where we share a tensor product state
$\rho^{\otimes n}$ which is the tensor product of 
a general state $\rho$ on the composite system 
$\cH_A \otimes \cH_B$.
We apply our protocol given in section 4 to 
the case where the entropy rate $H(\Tr_B \rho) $ is unknown.
In this situation, the coding length is variable,
the performance is characterized by 
the distribution of the coding length
${\rm P}^{{\bf E}^{\delta,n}}_{n,\rho}$
and the average of error 
$\epsilon_{n,\rho}({\bf E}^{\delta,n},{\bf D}^{\delta,n})$,
which are defined by
\begin{align*}
{\rm P}^{{\bf E}^{n}}_{n,\rho}(\omega_n)
& : =
\Tr {\bf E}^n_{\omega_n}(\rho^{\otimes n}) \\
\epsilon_{n,\rho}({\bf E}^{n},{\bf D}^{n})
& : =
\sum_{\omega}{\rm P}^{{\bf E}^{n}}_{n,\rho}(\omega_n)
\Tr {\bf E}^n_{\omega_n}(\rho^{\otimes n})
b^2\left( \rho^{\otimes n},
{\bf D}^n_{\omega_n} \otimes I_B^{\otimes n} 
\left(
\frac{
{\bf E}^n_{\omega_n} \otimes I_B^{\otimes n} 
( \rho^{\otimes n})}
{\Tr {\bf E}^n_{\omega_n}(\rho^{\otimes n})}
\right)
\right).
\end{align*}
As a general setting which unifies the above setting and the setting
given in section 3,
we consider the setting where 
a general state on the composite system $\cH_A \otimes \cH_B$ 
is generated with the probability $p(\rho)$.
In the i.i.d. extended setting,
a state $\vec{\rho}_n
:=\rho_1 \otimes \rho_2 \otimes \cdots \otimes\rho_n$
on the tensor product system 
$\cH_A^{\otimes n} \otimes \cH_B^{\otimes n}$ 
is generated with the i.i.d. probability $p^n$.
In this setting, the probability of the coding length 
${\rm P}^{{\bf E}^{\delta,n}}_{n,p}$
and the average of the error 
$\epsilon_{n,p}({\bf E}^{\delta,n},{\bf D}^{\delta,n})$
are defined by 
\begin{align*}
{\rm P}^{{\bf E}^{n}}_{n,p}(\omega_n)
& : =
\sum_{\vec{\rho}_n} p^n(\vec{\rho}_n)
\Tr {\bf E}^n_{\omega_n}(\vec{\rho}_n) \\
\epsilon_{n,p}({\bf E}^{n},{\bf D}^{n})
& : =
\sum_{\vec{\rho}_n} p^n(\vec{\rho}_n)
\sum_{\omega}{\rm P}^{{\bf E}^{n}}_{n,p}(\omega_n)
\Tr {\bf E}^n_{\omega_n}(\vec{\rho}_n)
b^2\left( \vec{\rho}_n,
{\bf D}^n_{\omega_n} \otimes I_B^{\otimes n} 
\left(
\frac{
{\bf E}^n_{\omega_n} \otimes I_B^{\otimes n} 
( \vec{\rho}_n)}
{\Tr {\bf E}^n_{\omega_n}(\vec{\rho}_n)}
\right)
\right).
\end{align*}
Since the case $\dim \cH_B=1$ is equivalent to
the setting in section 3 and 4,
this setting is a generalization of not only the above setting 
but also the setting in section 3 and 4.
Moreover, the encoder ${\bf E}^{\delta,n}$ and 
decoder ${\bf D}^{\delta,n}$ proposed in section 4
satisfies 
\begin{align}
\epsilon_{n,p}({\bf E}^{\delta,n},{\bf D}^{\delta,n})
& \le
1- \frac{[n(\delta-2 \delta')]}{[n \delta]}
\left(1- \left(n+d\right)^{4d}
\exp(-n C {\delta'}^2 )
\right)^{3/2} \label{19-3} \\
{\rm P}^{{\bf E}^{\delta,n}}_{n,p}
\left\{
\frac{1}{n}
\left( \log | \Omega_n |+ \log
\dim \cH_{k,i}\right)
\ge R
\right\}
& \le
(n+d)^{4d}
\exp\left(- n \min_{H({\bf b})\ge R - f(n,\delta)/n}
D({\bf b}\|{\bf a}) \right) \label{19-4} ,
\end{align}
where ${\bf a}$ is the probability distribution consists of 
the eigenvalues of $\overline{\rho}_{p,A} :=
\Tr_B \overline{\rho}_{p}$ and
$\delta'$ is arbitrary real number satisfying $0 \,<2 \delta' \,<
\delta$.
These inequalities are proven in Appendix,
Thus, we obtain two equations similar to (\ref{19-9}) and (\ref{19-10}).

\section{Discussion}
We construct a quantum universal variable-length code
for i.i.d. sources
from a quantum universal fixed-length code for i.i.d. sources.
This construction clarifies the relation between
the above types codes.
We can expect a similar relation in a more general setting, 
which is a future problem.

\section*{Acknowledgements}
The authors wish
to thank Professor H. Nagaoka and Dr. A. Winter 
for useful comments.
They are grateful to the referee for pointing out 
the possibility of the application to the compression of 
entangled states.

\appendix

First, we prove inequalities (\ref{9-5-1}) and (\ref{19-3}).
We can evaluate the average error as
\begin{align}
&\epsilon_{n,p}({\bf E}^{\delta,n},{\bf D}^{\delta,n}) 
\nonumber \\
= & \sum_{\vec{\rho}_n}
p^n(\vec{\rho}_n)
\sum_{k,i}
\Tr \left[
\frac{1}{[n\delta]}
\left(P^{\vec{\alpha}(k/n),n}_i \otimes I_B^{\otimes n} \right)
\vec{\rho}_n\right] \nonumber \\
& \times \left( 1- \Tr \left|\sqrt{\vec{\rho}_n}
\sqrt{\frac{\sqrt{
\frac{1}{[n\delta]}
\left(P^{\vec{\alpha}(k/n),n}_i \otimes I_B^{\otimes n} \right)
}
\vec{\rho}_n
\sqrt{\frac{1}{[n\delta]}
\left(P^{\vec{\alpha}(k/n),n}_i \otimes I_B^{\otimes n} \right)
}}
{\Tr  
\frac{1}{[n\delta]}
\left(P^{\vec{\alpha}(k/n),n}_i \otimes I_B^{\otimes n} \right)
\vec{\rho}_n}
}\right|\right) \nonumber \\
 = &
1- \sum_{\vec{\rho}_n }
p^n(\vec{\rho}_n)
\sum_{k,i}
\sqrt{\Tr 
\frac{1}{[n\delta]}
\left(P^{\vec{\alpha}(k/n),n}_i\otimes I_B^{\otimes n} \right)
\vec{\rho}_n}\nonumber \\
& \times 
\Tr \sqrt{\sqrt{\vec{\rho}_n}
\sqrt{
\frac{1}{[n\delta]}
\left(P^{\vec{\alpha}(k/n),n}_i\otimes I_B^{\otimes n} \right)
}
\vec{\rho}_n
\sqrt{
\frac{1}{[n\delta]}
\left(
P^{\vec{\alpha}(k/n),n}_i\otimes I_B^{\otimes n} \right)
}
\sqrt{\vec{\rho}_n}} \nonumber\\
 = &
1- 
\sum_{k,i}
\frac{1}{[n \delta]}
\sum_{\vec{\rho}_n }
p^n(\vec{\rho}_n)
\left(\Tr 
\left(P^{\vec{\alpha}(k/n),n}_i\otimes I_B^{\otimes n} \right)
\vec{\rho}_n
\right)^{\frac{3}{2}}
\nonumber\\
 \le &
1- 
\sum_{k,i}
\frac{1}{[n \delta]}
\left(
\sum_{\vec{\rho}_n }
p^n(\vec{\rho}_n)\Tr 
\left(P^{\vec{\alpha}(k/n),n}_i\otimes I_B^{\otimes n} \right)
\vec{\rho}_n
\right)^{\frac{3}{2}} \label{jensen} \\
 = &
1- 
\sum_{k,i}
\frac{1}{[n \delta]}
\left( \Tr \overline{\rho}_p^{\otimes n}
\left(P^{\vec{\alpha}(k/n),n}_i\otimes I_B^{\otimes n} \right)
\right)^{\frac{3}{2}} \nonumber \\
 = &
1- 
\sum_{k,i}
\frac{1}{[n \delta]}
\left( \Tr_A \overline{\rho}_{p,A}^{\otimes n}
P^{\vec{\alpha}(k/n),n}_i
\right)^{\frac{3}{2}} ,\label{e51}
\end{align}
where inequality (\ref{jensen}) follows from 
Jensen's inequality concerning 
the convex function $x \mapsto x^{3/2}$.
Note that in the case of inequality (\ref{9-5-1}), 
$\dim \cH_B =1$ and $ \overline{\rho}_{p,A}^{\otimes n}
=  \overline{\rho}_{p}^{\otimes n}$.
The number of the pair $(k,i)$ satisfying 
$|H({\bf a}) - \alpha(k/n)_i- \frac{\delta}{2}| 
\le \frac{\delta}{2}- \delta'$
is $[n(\delta-2 \delta')]$ or $[n(\delta-2 \delta')]+1 $.
When the pair $(k,i)$ satisfies this condition,
$\alpha(k/n)_i \le H({\bf a}) - \delta'$
and $\alpha(k/n)_{i+1} \ge H({\bf a}) + \delta'$.
Therefore, using (\ref{10-2}) and (\ref{103}) we obtain
\begin{align}
\Tr \overline{\rho}_{p,A}^{\otimes n}
P^{\vec{\alpha}(k/n),n}_i
\ge
1- \left(n+d\right)^{4d}
\exp(-n C \delta^{'2} ).
\label{910}
\end{align}
Inequalities (\ref{9-5-1}) and (\ref{19-3}) follow from 
(\ref{910}) and (\ref{e51}).

Next, we prove (\ref{9-5-2}) and (\ref{19-3}).
Inequality (\ref{10-1}) guarantees that
\begin{align*}
\dim \cH_{k,i} \le \rank P_{\alpha(k/n)_{i+1},n}
\le (n+d)^{4d} e^{n \alpha(k/n)_{i+1}}.
\end{align*}
For any $k$, we let $i_k$ be the minimum integer satisfying
\begin{align}
\frac{1}{n}
\left( \log | \Omega_n |+ \log
\dim \cH_{k,i}\right)
\ge R,  \label{10-3}
\end{align}
i.e., 
$\dim \cH_{k,i} \ge e^{nR}/ ([n \delta] l)$.
Since $l = (\log d)/ \delta +1$, 
\begin{align}
 \alpha(k/n)_{i_k} \ge
R+ f(n,\delta)/n. \label{10-4}
\end{align}
From (\ref{10-2}) and (\ref{10-4}),
we obtain
the inequality 
\begin{align*}
\sum_{i: (\ref{10-3})}
\Tr P_{i}^{\vec{\alpha}(k/n),n} \overline{\rho}_{p,A}^{\otimes n}
=
\Tr P_{\alpha(k/n)_{i_k},n} \overline{\rho}_{p,A}^{\otimes n}
\le
(n+d)^{4d}
\exp\left(- n \min_{H({\bf b})\ge
R+ f(n,\delta)/n}
D({\bf b}\|{\bf a}) \right).
\end{align*}
Thus,
\begin{align*} 
\frac{1}{[n \delta]}\sum_k \sum_{i: (\ref{10-3})}
\Tr P_{i}^{\vec{\alpha}(k/n),n} \overline{\rho}_{p,A}^{\otimes n}
\le
(n+d)^{4d}
\exp\left(- n \min_{H({\bf b})\ge
R+ f(n,\delta)/n}
D({\bf b}\|{\bf a}) \right),
\end{align*}
which implies (\ref{9-5-2}) and (\ref{19-3}).

Finally, we prove (\ref{19-7}).
Since $P_{R,n} \vec{\rho}_n P_{R,n}
+ \Tr [(I-P_{R,n}) \vec{\rho}_n 
\frac{P_{R,n}}{\Tr P_{R,n}}
\ge P_{R,n} \vec{\rho}_n P_{R,n} $ and 
the function $x \mapsto \sqrt{x}$ is operator monotone,
we obtain $\sqrt{P_{R,n} \vec{\rho}_n P_{R,n} 
+ \Tr [(I-P_{R,n})\vec{\rho}_n]
\frac{P_{R,n}}{\Tr P_{R,n}}}
\ge
\sqrt{P_{R,n} \vec{\rho}_n P_{R,n} }
$.
Therefore,
\begin{align}
&  \epsilon (E^n_{R},D^n_{R})
\nonumber \\
 = &
\sum_{\vec{\rho}_n} p^n(\vec{\rho}_n)
\left[
1- \left(\Tr \left|
\sqrt{P_{R,n} \vec{\rho}_n P_{R,n} 
+ \Tr [(I-P_{R,n})\vec{\rho}_n]
\frac{P_{R,n}}{\Tr P_{R,n}}}
\sqrt{\vec{\rho}_n} 
\right|
\right)^2
\right] \nonumber\\
 \le &
\sum_{\vec{\rho}_n} p^n(\vec{\rho}_n)
\left[
1- \left(\Tr \left|
\sqrt{P_{R,n} \vec{\rho}_n P_{R,n} }\sqrt{\vec{\rho}_n} 
\right|
\right)^2
\right] \nonumber\\
 = &
\sum_{\vec{\rho}_n} p^n(\vec{\rho}_n)
\left[
1- \left(\Tr 
\sqrt{\sqrt{\vec{\rho}_n} P_{R,n} \vec{\rho}_n P_{R,n} \sqrt{\vec{\rho}_n} }
\right)^2
\right]  \nonumber\\
=& 
\sum_{\vec{\rho}_n} p^n(\vec{\rho}_n)
\left[
1- \left(\Tr \sqrt{\vec{\rho}_n} 
P_{R,n} \sqrt{\vec{\rho}_n} 
\right)^2
\right] \nonumber\\
 = &
\sum_{\vec{\rho}_n} p^n(\vec{\rho}_n)
\left[
1- \left(\Tr \vec{\rho}_n P_{R,n} 
\right)^2
\right] \nonumber \\
\le &
\sum_{\vec{\rho}_n} p^n(\vec{\rho}_n)
2 \left[
1- \Tr \vec{\rho}_n P_{R,n} 
\right] \label{8-8} \\
 = &
2 \left(
1- \Tr \left[\sum_{\vec{\rho}_n} p^n(\vec{\rho}_n)\vec{\rho}_n P_{R,n} 
\right]
\right) 
=
2 \left(
1- \Tr \overline{\rho}_p^{\otimes n} P_{R,n} 
\right) , \nonumber 
\end{align}
where inequality (\ref{8-8}) follows from 
the inequality $1-x^2 \le 2(1-x)$.
Therefore, using inequality (\ref{10-2}), we obtain (\ref{19-7}).

\end{document}